\documentclass[twocolumn,osajnl,showpacs,bibnotes,floatfix]{revtex4}

\usepackage{graphics}

\newcommand{\ket}{\rangle}
\newcommand{\bra}{\langle}
\newcommand{\op}{\widehat}
\newcommand{\iin}{_{\rm in}}
\newcommand{\iout}{_{\rm out}}
\newcommand{\iest}{_{\rm est}}

\begin{document}

\title{Image processing as state reconstruction in optics}

\author{M. Je\v{z}ek}
\email{jezek@optics.upol.cz}
\author{Z. Hradil}

\affiliation{Department of Optics, Palack\'{y} University,
  17. listopadu 50, 77200 Olomouc, Czech Republic}
\date{\today}


\begin{abstract}
The image reconstruction of partially coherent light is
interpreted as the quantum state reconstruction.
The efficient method based on maximum-likelihood
estimation is proposed to acquire information from registered
intensity measurements affected by noise.
The connection with totally incoherent image
restoration is pointed out.
The feasibility of the method is demonstrated numerically.
Spatial and correlation details significantly smaller than
the diffraction limit are revealed in the reconstructed pattern.
\end{abstract}


\maketitle

\section{Introduction}

Light conveys considerable part of information on the
surrounding world. Information coded into and transmitted
by means of light plays the key role in contemporary
information technology. That is why any deeper understanding
of fundamental limitations imposed by the theory represents
a challenging problem.
The origin of the {\em diffraction limit} restricting the spatial
resolution is comprehended since the era of wave optics.
This phenomenon, manifested for example by a fuzzy diffraction
spot as a consequence of
the finite aperture, yields a serious limitation in image
processing.
But there are still other physical restrictions
which must be taken into account. Direct observations are
not able to determine the phase of optical fields due to the
fast oscillations and due to the effect of the time integration
of intensity detectors. The phase must be therefore retrieved
adopting sophisticated techniques. This is known
{\em phase problem\/}, solution of which is sensitive to noise
and requires various regularization treatments.

There are several ways to surpass the mentioned
limitations imposed by realistic aspects of optical observations.
Sophisticated algorithms of data processing and image
reconstruction have been devised, such as analytic continuation
of signal and diagonalization of optical device
\cite{Burge_Fiddy_Greenaway_Ross_1974,%
Burge_Fiddy_Greenaway_Ross_1976,%
Frieden_1967,Frieden_1969,Francia_1969,Frieden_in_Progress,%
Perina_Perinova_1969,Perina_Perinova_Braunerova_1977,%
Perina_Coherence_of_Light},
different regularization and smoothing techniques
of direct deconvolution and
methods of projections onto convex sets (POCS)
\cite{Phillips_1962,Twomey_1963,%
Miller_1970,Tikhonov_Arsenin,%
Backus_Gilbert_1968,Backus_Gilbert_1970,%
Gerchberg_Saxton_1972,%
Gerchberg_1974,Fienup_1982,Kim_2001,Gerchberg_2002},
solving of the transport-of-intensity equation
for phase reconstruction
\cite{Teague_1982,Teague_1983,%
Nugent_et_al_1996,Paganin_Nugent_1998,%
Allen_Oxley_2001},
utilizing of canonical transforms
\cite{Bastiaans_Wolf_2003},
and statistical methods based on
minimum least-squares distance
\cite{Helstrom_1967,Biraud_1969},
maximum entropy
\cite{Frieden_1972,Gull_Daniell_1978,Frieden_1998},
maximum Cramer-Rao bound
\cite{Frieden_1988,Rehacek_Hradil_2002_INF},
and maximum likelihood principle
\cite{Richardson_1972,%
Rockmore_Macovski_1976,Rockmore_Macovski_1977,%
Dempster_Laird_Rubin_1977,Shepp_Vardi_1982,%
Snyder_at_al_1987,Vardi_Lee_1993,Rehacek_et_al_2002}.
The tomographical synthesis of different
intensity observations of an object can considerably
improve its reconstruction and allow the phase retrieval
\cite{Bertrand_Bertrand_1987,Vogel_Risken_1989,
Smithey_Beck_Raymer_Faridani_1993,Raymer_Beck_McAlister_1994,%
D'Ariano_Macchiavello_Paris_1994,James_Agarwal_1995,%
Mancini_Man'ko_Tombesi_1995,%
Banaszek_Wodkiewicz_1996,Schiller_Breitenbach_et_al_1996,%
Breitenbach_Schiller_Mlynek_1997,Kurtsiefer_Pfau_Mlynek_1997,%
Lvovsky_Hansen_et_al_2001}.
The resolution can also be enhanced by eliminating of
out-of-focus light by means of the apodisation technique
\cite{Apodisation}
or by utilizing of confocal arrangement and interaction
between light and matter, like in multi-photon fluorescence
microscopy and STED technique
\cite{Dyba_Hell_2002,Stelzer_2002}.

In this article the problem of image processing will be
addressed from the viewpoint of statistical reconstruction
techniques. The series of tomographical-like intensity measurements
will be used up for the reconstruction of a state of partially
coherent light. In the particular case of totally incoherent
light the proposed approach will be identified with
the Richardson algorithm \cite{Richardson_1972} of the
image reconstruction.

There is a tight connection between fundamental principles
of wave optics and quantum mechanics. Description of the scalar
wave in optics is equivalent to description of the de~Broglie
wave of a mass particle in the framework of quantum mechanics.
The pure quantum state in position representation coincides with
the complex scalar wave, and similarly, the mixed quantum state
in this representation corresponds to the correlation function
of the second order. This connection will be systematically exploited
and the problem of quantum state reconstruction will be considered
in analogy with optical counterpart of image processing. 

For the sake of simplicity all the considered problems will
be treated as two dimensional problems. The first dimension
corresponds to the evolution parameter---time $t$ in dynamical
problems or longitudinal $z$-coordinate in the case of image
processing. The second dimension corresponds to the observed
quantity. This could be position in the former case of dynamical
problems or transverse $x$-coordinate in the later case  of image
processing. Further generalization to higher dimension can be
obtained by a straightforward manner.

\section{Wave theory}

In this section the analogy between scalar wave optics and quantum
mechanics will be highlighted. As will be shown, abstract quantum
formulation is advantageous for the purpose of signal
reconstruction.

In quantum domain, pure quantum state $|\psi\rangle$ from the
Hilbert space represents the complete knowledge about the position
and momentum of a particle, of course obeying the Heisenberg
uncertainty principle. Any randomness in the ensemble of
identically prepared particles is described by the incoherent mixture
of pure states---a density operator,
\begin{equation} \label{density_operator}
  \op{\rho} = \sum_{k} \lambda_{k} |\varphi_k\ket \bra\varphi_k|.
\end{equation}
Probabilities $\lambda_{k}$ are nonnegative and adds to unity,
and the {\em mixed state} (\ref{density_operator}) satisfies
the following relations,
\begin{equation} \label{density_operator_properties}
  \op{\rho}^{+} = \op{\rho}, \quad
  {\rm Tr}[\op{\rho}] = 1, \quad
  \bra\psi|\op{\rho}|\psi\ket \geq 0, \,\, \forall |\psi\ket,
\end{equation}
where ${}^{+}$ means the Hermitian conjugation.
Denoting formally the position by a projector $|x\ket\bra x|$,
complex amplitude $\psi(x) = \bra x|\psi\ket$ describes
the coherent quasi-monochromatic 
scalar field.
Indeed, it characterizes the amplitude as well as the
phase of the propagating wave.
In the general case of partially coherent field, the
second order correlation function \cite{Perina_Coherence_of_Light}
\begin{eqnarray}  \label{mutual_intensity}
 \Gamma(x,x') & = &
   {\rm Tr}\left[ \op{\rho} |x'\ket\bra x| \right] =
   \sum_k \lambda_k
   \bra x|\varphi_k\ket \bra\varphi_k|x'\ket = \nonumber \\
   & = &
   \sum_k \lambda_k \varphi_k(x) \varphi_k^{\ast}(x') =
   \bra \psi^{\ast}(x') \psi(x) \ket_{\rm ens},
\end{eqnarray}
describes the statistical properties of the field.
The brackets $\bra~\ket_{\rm ens}$ denote the averaging
over complex amplitudes of all modes $k$.
The analogy between the density matrix (\ref{density_operator})
and the {\em mutual intensity} (\ref{mutual_intensity})
is expressed clearly by the relations analogous to
(\ref{density_operator_properties}),
\begin{equation}  \label{mutual_intensity_properties}
 \Gamma^{\ast}(x',x) = \Gamma(x,x'), \quad
 \int\!{\rm d}x\, I(x) = 1, \quad
 \Gamma(x,x) \ge 0.
\end{equation}
Notice that the function $I(x)=\Gamma(x,x)$ means the
optical intensity of field.
The analogy between quantum and wave descriptions
can be emphasized in phase space by means of the
Wigner quasi-distribution \cite{Wigner_1932,Ville_1948},
\begin{equation}  \label{Wigner_distribution}
  W(x,p) = \frac{1}{\pi} \int \! {\rm d}x' \,
  e^{-{\rm i} 2 p x'} \, \Gamma(x+x',x-x').
\end{equation}
This $(x,p)$ distribution is real bounded function,
which is however, not positively defined in general.

Let us proceed further to consider the state transformation.
Assuming linearity and causality the equation for evolution
of pure state formally reads
\begin{equation}  \label{pure_state_evolution}
  |\psi\ket\iout = \op{T} |\psi\ket\iin.
\end{equation}
Here $\op{T}$ is linear operator satisfying the equation
\begin{equation}  \label{evolution_equation}
  \frac{\partial}{\partial z} \op{T} = \op{L} \op{T},
\end{equation}
where $z$ is evolution parameter and the generator $\op{L}$
of evolution is considered to be $z$-independent.
The evolution equation (\ref{evolution_equation})
covers both the Schr{\"{o}}dinger equation in quantum mechanics
and paraxial Helmholtz equation in Fresnel approximation of
scalar wave optics. The {\em unitary evolution\/} is governed
by the Hamiltonian operator,
$\op{L} = -\frac{\rm i}{\hbar} \op{H}$,
and the evolution of the mixed state is described by
transformation
\begin{equation}  \label{density_operator_evolution}
  \op{\rho}\iout = \op{T} \, \op{\rho}\iin \, \op{T}^{+},
  \qquad
  \op{T} =  \exp{\left[ -\frac{\rm i}{\hbar} \op{H} z \right]}.
\end{equation}

Evolution (\ref{density_operator_evolution}) of the state
$\op{\rho}$ in the Schr{\"{o}}dinger picture can be equivalently
formulated in the Heisenberg picture. This formulation
follows the laws of ray optics.
Indeed, the relationship between the canonical
observables of {\em position\/} and {\em momentum\/} reads
\begin{equation}  \label{x_p_operators_evolution}
  \left( \matrix{\op{x}\iout \cr \op{p}\iout} \right) =   
  \op{T}^{+} \left( \matrix{\op{x}\iin \cr \op{p}\iin} \right)
  \op{T}.
\end{equation}
Particularly, for the evolution generated by quadratic Hamiltonian
in canonical observables the transformation
(\ref{x_p_operators_evolution}) is linear,
\begin{equation}  \label{linear_x_p_operators_evolution}
  \left( \matrix{\op{x}\iout \cr \op{p}\iout} \right) =
  T \left( \matrix{\op{x}\iin \cr \op{p}\iin} \right) =
  \left( \matrix{ A & B \cr C & D } \right)
  \left( \matrix{\op{x}\iin + s \cr \op{p}\iin + r} \right),
\end{equation}
where $\mbox{Det}[T] = AD - BC = 1$.
This generic $ABCD$ transformation covers useful cases of wave
transformation, for example free evolution, displacement,
rotation, phase shift, squeezing, and chirp.
Linear transformation of the $(\op{x},\op{p})$ operators
induces the evolution of the Wigner function by linear
transformation of its variables,
\begin{equation}  \label{linear_transformation_of_Wigner_function}
  W(x,p) = W(Dx-Bp-s,-Cx+Ap-r).
\end{equation}
Roughly speaking, all these transformations rotate and rescale
the input state. As the consequence, the observable
$A\op{x}+B\op{p}$ can be measured offering an important tool
for all the tomographical methods.

In the {\em classical limit\/} there is a tight connection between
evolution of position and momentum operators in the Heisenberg
picture
(\ref{linear_x_p_operators_evolution}) and geometrical paraxial
optics represented by the identity between operators
$(\op{x},\op{p})$ and its c-values $(x,p)$. The state vector
$(x,p)$ is used to specify position and angle of the optical
ray. Similar description may be adopted for particle
in classical mechanics. Evolution operator $\op{T}$ is given by
the $ABCD$ matrix $T$ completed by the transverse shift $s$
and rotation $r$ in analogy with the relation
(\ref{linear_x_p_operators_evolution}).
Geometrical optics as well as classical mechanics do not involve
interference, what simplifies the input-output relations considerably.
This is why the geometrical optics is so suitable for
``direct'' observations. Indeed, if  the positions $x_{1}$, $x_{2}$
for the two values $z_{1}$, $z_{2}$ are measured,  the state vector
$(x_{0},p_{0})$ for $z=0$ can be completely reconstructed as
\begin{equation}
  \left( \matrix{x_{0} \cr p_{0}} \right) = \frac{1}{z_{2}-z_{1}}
  \left( \matrix{x_{1}z_{2}-x_{2}z_{1} \cr x_{2}-x_{1}} \right).
\end{equation}

In the case of losses the evolution turns out to be {\em non-unitary\/}.
Let us imagine the absorbing screen with $2a$ aperture.
The incident state can be decomposed in the base of eigenstates
$|\xi\ket$ of position $\op{x}$ on screen.
Since only a part of eigenstates spectrum for eigenvalues
$\xi \in [-a,a]$ is transmitted, the non-unitary transformation
can be described by projection operator
\begin{equation}  \label{finite_aperture_transformation_operator}
  \op{T} = \int_{-a}^{a} \! {\rm d}\xi \, |\xi\ket \bra\xi|
\end{equation}
corresponding to finite aperture.

Let us conclude the overview by explicit formulation
of the state transformation in position representation.
The generic evolution (\ref{pure_state_evolution}) of the pure
state (coherent wave) takes a well-known form
of the {\em superposition integral\/},
\begin{eqnarray}
  \psi\iout(x) & = & \bra x |\psi \ket\iout =
  \int \!{\rm d}x_{0}\, \bra x|\op{T}|x_{0} \ket
       \bra x_{0}|\psi \ket\iin = \nonumber \\
  & = &
  \int \!{\rm d}x_{0}\, h(x,x_{0}) \, \psi\iin(x_{0}).
  \label{superposition_integral_for_pure_state}
\end{eqnarray}
The kernel of the integral transformation
(\ref{superposition_integral_for_pure_state}),
\begin{equation}  \label{PSF}
  h(x,x_{0}) = \bra x| \op{T} |x_{0}\ket,
\end{equation}
is called propagator in the quantum theory and the response
function or {\em point-spread function\/} (PSF) in the scalar
wave theory \cite{Goodman_Fourier_Optics}.
Loosely speaking, it relates the point source
in the object (input) plane, $z=0$, with its image in the image
(output) plane with coordinate $z$. This mapping is fuzzy in
realistic image processing due to the effect of diffraction
caused by non-unitary evolution.
In the case of unlimited aperture it may become sharp
corresponding to the case of unitary evolution,
$\op{T}^{+}=\op{T}^{-1}$.
Analogously, the evolution (\ref{density_operator_evolution})
of the mixed state (mutual intensity) in the position
representation reads
\begin{equation}  \label{superposition_integral_for_mixed_state}
  \Gamma\iout(x,x') = \int \!\!\! \int \!{\rm d}q{\rm d}q'\,
  h(x,q) h^{\ast}(x',q') \, \Gamma\iin(q,q').
\end{equation}

\section{Detection}

According to standard formulation of quantum mechanics the
measurement is represented by an observable,
a Hermitian operator $\op{A}$.
Eigenvalues of this operator correspond to possible results
of elementary measurements. Eigenstates determine the possible
states after the measurement and they are complete and orthogonal,
\begin{equation}
  \op{A} |a\ket = a |a\ket,  \quad
  \sum_{a} |a\ket \bra a| = \op{1},  \quad
  \bra a | a'\ket = \delta_{aa'}.
\end{equation}
These properties are reflected in the probability
$p_{a} = {\rm Tr}[\op{\rho} \, |a\ket \bra a|]$
predicted by quantum theory guaranteeing the normalization
of probabilities $\sum_{a} p_{a} = 1$ (completeness),
and mutual exclusivity of the results $a$ (orthogonality).
This description may be further generalized in terms of
positive-operator valued measure (POVM) yielding
the decomposition of identity operator \cite{Holevo_1973,Helstrom_QDET},
\begin{equation}
  \op{\Pi}_{b} \ge 0, \qquad  \sum_{b} \op{\Pi}_{b} = \op{1}.
\end{equation}
It predicts the probability for registering the output $b$
analogously to the case of orthogonal projectors,
$p_{b} = {\rm Tr}[\op{\rho} \, \op{\Pi}_{b}]$.
The notion of POVM plays the crucial role in description
a generic quantum measurement in state estimation and
discrimination.

Registration of the image intensity $I(x)$
of partially coherent wave in the transverse position $x$
corresponds to the measurement of position operator
$\op{x}$ in the output plane,
\begin{equation}  \label{optical_intensity}
  I(x) = \Gamma\iout(x,x) = p(x) =
  {\rm Tr}\left[ \op{\rho}\iout |x \ket\bra x| \right].
\end{equation}
Realistic detector always possesses the finite spatial resolving
power. Denoting its pixels by the indices $i$, the simplest
representation of detector POVM is given by the operators
\begin{equation}  \label{pixel_projector}
  \op{O}_{i} =  \int_{\Delta_{i}} \! {\rm d}x \, |x\ket \bra x|,
\end{equation}
where the integration is done along the surface of the $i$-th
pixel.

Consider now the generic observation scheme. The input
state $\op{\rho}$ represented by its mutual intensity
$\Gamma(x,x')$ in wave description is transformed by
optical device $\op{T}=\op{T}(A,B,\ldots)$ with the response
function $h(x,x_{0};A,B,\ldots)$.
Resulting output state $\op{\rho}\iout$ is observed by the
detector (\ref{pixel_projector}) placed in the output plane.
The detector counts the elementary clicks in every $i$-th pixel.
The numbers $N_{i}$ of registered clicks represented by
relative frequencies $f_{i} = N_{i}/N$, $N=\sum_{i}N_{i}$,
sample the probabilities $p_i$ (intensities $I_i$),
\begin{equation}  \label{flash_in_pixel_probability}
  p_{i} = {\rm Tr} \left[ \op{\rho}\iout \op{O}_{i} \right] =
  {\rm Tr} \left[ \op{\rho} \, \op{\Pi}_i \right], \quad
  \op{\Pi}_{i} = \op{T}^{+} \op{O}_{i} \op{T}.
\end{equation}
Notice however, that this scheme is rather classical and it does
not take into account statistics of detection process in
accordance with classical image processing, when intensity
is considered as measurable quantity.
Provided that quantum nature of detection will be considered,
the relation (\ref{pixel_projector}) should be modified
taking into account registration of photons instead.

\section{Direct reconstruction}

The reconstruction of the signal in wave theory is rather
involved and extensive field with many applications.
Let us review briefly this topic. Assuming
an unknown signal propagating through optical refractive and
diffractive elements, the output field may be detected.
Provided that properties of the optical device are known,
and detection is ideal, the input signal may be predicted
from the output one. This is the classical inverse problem
of wave optics.

Standard methods use the isoplanatic approximation involving
the relation (\ref{superposition_integral_for_pure_state})
as convolution,
\begin{equation}
  \psi\iout = \int \! {\rm d}x_{0} \,
  h(x-x_{0}) \, \psi\iin(x_{0}) =
  h \, \ast \, \psi\iin.
\end{equation}
Inversion is given by the {\em Fourier deconvolution}
\begin{equation}  \label{Fourier_deconvolution}
  \widetilde{\psi}\iin =
  \frac{\widetilde{\psi}\iout + \widetilde{\cal N}}{\widetilde{h}}.
\end{equation}
Here $\widetilde{\psi}\iin$, $\widetilde{\psi}\iout$, and
$\widetilde{h}$ are Fourier transformations
of $\psi\iin$, $\psi\iout$, and ${h}$, respectively,
and $\widetilde{\cal N}$ represents the spectrum of
additive noise ${\cal N}$. Typical
point-spread function ${h}$ has the form of Sinc or BeSinc
function and $\widetilde{h}$ corresponds to step-function.
Hence, the spatial frequencies of the signal are transmitted
only up to certain upper cut-off \cite{Goodman_Fourier_Optics}.
This is why the deconvolution is very sensitive to noise
${\cal N}$ and diverges at spatial frequencies for which
the transfer function $\widetilde{h}$ turns to be zero.
In particular, for frequencies
above the cut-off, the transfer function $\widetilde{h}$
vanishes and (\ref{Fourier_deconvolution}) diverges
due to the broad noise spectrum $\widetilde{\cal N}$.
Some regularization procedures are necessary
in all these cases \cite{Phillips_1962,Twomey_1963,%
Miller_1970,Tikhonov_Arsenin,%
Backus_Gilbert_1968,Backus_Gilbert_1970,%
Gerchberg_Saxton_1972}.
The special attention has been devoted to the more
accurate description of the optical device. Detailed
analysis needs special choice of eigenfunctions related 
to finite aperture instead of spatial frequencies
\cite{Frieden_1967,Frieden_1969,Francia_1969}.
Systematic theory of this remarkable basis, so called 
prolate spheroidal wave functions, was given by
Frieden\cite{Frieden_in_Progress}.
Several further super-resolution techniques as 
apodisation \cite{Apodisation} or analytic continuation 
\cite{Perina_Perinova_1969,Perina_Perinova_Braunerova_1977}
have been suggested.
The inverse source problem may be further generalized,
taking into account other realistic aspects of detection.
For example, optical intensity is detected by real
photo-detectors instead of complex amplitude and phase
is subject of reconstruction
\cite{Burge_Fiddy_Greenaway_Ross_1974,%
Burge_Fiddy_Greenaway_Ross_1976,Perina_Coherence_of_Light,%
Gerchberg_Saxton_1972,%
Fienup_1982,Kim_2001,Gerchberg_2002,%
Teague_1982,Teague_1983,%
Nugent_et_al_1996,Paganin_Nugent_1998,%
Allen_Oxley_2001,Bastiaans_Wolf_2003}.

Standard image processing deals with the observation in the
image plane revealing the sharpest image. However,
the observations in defocused planes are also worthwhile
\cite{Raymer_Beck_McAlister_1994,%
James_Agarwal_1995,Kurtsiefer_Pfau_Mlynek_1997,%
Liu_Brenner_2003}.
They correspond to the observation of $A\op{x} + B\op{p}$
operators in the language of quantum
theory (\ref{linear_x_p_operators_evolution}),
bringing other piece of information about signal and
affording better employment of measured data.
Such tomographical technique was suggested by Bertrand,
Bertrand \cite{Bertrand_Bertrand_1987} and Vogel, Risken
\cite{Vogel_Risken_1989} and
experimentally verified by group from University of Oregon
\cite{Smithey_Beck_Raymer_Faridani_1993}.
Up to now several other tomographical schemes for observing
of various faces of the system have been proposed.
This is usually achieved by adjusting of some parameters
in the set-up. Particular configuration in dependence
on parameter provides the desired group of transformations.
Both the classical X--ray tomography (CT) used in medicine
\cite{Herman_CT,Kak_Slaney_CT,Natterer_CT} and
the homodyne tomography \cite{Vogel_Risken_1989,%
Smithey_Beck_Raymer_Faridani_1993,Schiller_Breitenbach_et_al_1996,%
Breitenbach_Schiller_Mlynek_1997,Lvovsky_Hansen_et_al_2001}
use the group of rotations.
Phase-space tomography and chronocyclic tomography are connected
with the symplectic group but they are convertible to the classical
tomography \cite{Raymer_Beck_McAlister_1994,James_Agarwal_1995,%
Kurtsiefer_Pfau_Mlynek_1997}.
General non-homogeneous symplectic tomography was introduced
by Mancini, Man'ko, and Tombesi \cite{Mancini_Man'ko_Tombesi_1995}.

All the above mentioned inverse problems are relating measured
data $f_i$ with theoretical probabilities $p_i$ by means
of the equality
\begin{equation}  \label{p=f}
  {\rm Tr}\left[ \op{\rho} \, \op{\Pi}_i \right] = f_{i},
\end{equation}
where multi-index $i$ passes over all configurations of optical
device and over all pixels of detector. However, the solution
of linear equations (\ref{p=f}) represents an ill-posed problem
of the same kind as image reconstruction using deconvolution.
This procedure is very sensitive to noise, which is
an inevitably involved in any measurement scheme.
Ill posed problem implies, that reconstructed ``state'' need
not represent any physically possible object. In the language
of quantum theory this means that
$\bra\psi|\op{\rho}|\psi\ket<0$ may hold for some states.
Alternatively, the optical intensity $I(x)$ may drop below
zero at some position coordinates in wave-theory language.
Linear algorithm is not capable to guarantee such necessary
conditions as positive definiteness of density matrix
or mutual intensity.

Statistical approaches suggest a remedy to this problem
releasing the strict condition (\ref{p=f}).
For example, the equality between $f_i$ and $p_i$ could be
replaced by requirement of their minimal least-squares
distance
\begin{equation}  \label{least_squares}
  \sum_i \left|
    f_i - {\rm Tr}\left[ \op{\rho} \, \op{\Pi}_i \right]
  \right|^2,
\end{equation}
an obvious choice in engineering practice.
But the other metrics are also eligible.
Least-squares method
\cite{Helstrom_1967,Biraud_1969,Opatrny_Welsch_Vogel_1997},
Richardson method \cite{Richardson_1972},
maximum-likelihood (ML) principle
and expectation-maximization (EM) algorithm
\cite{Fisher_1922,Fisher_1925,%
Rockmore_Macovski_1976,Rockmore_Macovski_1977,%
Dempster_Laird_Rubin_1977,Shepp_Vardi_1982,%
Snyder_at_al_1987,Vardi_Lee_1993,Rehacek_et_al_2002},
principle of maximum Cramer-Rao bound
\cite{Frieden_1988,Rehacek_Hradil_2002_INF},
maximum entropy (ME) method
\cite{Shannon_1948,Jaynes_1957,Frieden_1972,%
Gull_Daniell_1978,Frieden_1998,Buzek_Adam_Drobny_1996},
and intrinsic correlation function (ICF) model represent
several examples of various statistical signal-processing
schemes. In the following section the arguments in favor
of maximum likelihood estimation will be formulated with
the help of quantum treatment.

\section{Reconstruction as generalized quantum measurement}

Let us assume the generic scheme for the quantum measurement
(\ref{pixel_projector})-(\ref{flash_in_pixel_probability})
described above.
The conditional probability of detecting $N_{i}$ clicks
in the $i$-th pixel, if the state $\op{\rho}$ occurs in the
input plane, has the form of multinomial distribution
\begin{equation}  \label{multinomial_distribution}
  {\mathcal L}(\op{\rho}) \approx \prod_{i} p_{i}^{N f_i},
\end{equation}
where $f_i$ are the relative frequencies of registered
clicks, $N f_i = N_i$.
The input state $\op{\rho}$ is the subject of estimation
procedure. The {\em likelihood functional}
(\ref{multinomial_distribution}) then gives the answer
to the question ``How is it likely that the given data
$f_i$ were registered provided that the system was in
the given quantum state $\op{\rho}$\/?''
For some states the detection of given data is more
likely than for others.
Using the relation (\ref{flash_in_pixel_probability})
the log-likelihood function reads
\begin{equation}  \label{log_likelihood}
  \ln{\mathcal L}(\op{\rho}) =
  \sum_{i} f_{i} \ln p_{i} =
  \sum_{i} f_{i} \ln {\rm Tr}\left[ \op{\rho}\,\op{\Pi}_i \right].
\end{equation}
{\em Maximum likelihood\/} principle selects such a state
$\op{\rho}\iest$ for which the likelihood reaches its maximum,
\begin{equation}  \label{maximum_log_likelihood}
  \op{\rho}\iest = {\rm arg}\left[ \max_{\hat\rho} \,
  \ln {\mathcal L}(\op{\rho}) \right].
\end{equation}
The formal necessary condition
\begin{equation}  \label{variation_of_log_likelihood}
  \left. \frac{\delta \ln{\mathcal L}(\op{\rho})}
  {\delta \op{\rho}} \right|_{\hat\rho\iest} = 0
\end{equation}
may be rewritten to the form of {\em extremal operator equation\/}
\cite{Hradil_1997,Hradil_Summhammer_Rauch_1999,%
Hradil_Summhammer_2000,Rehacek_Hradil_Jezek_2001,%
Jezek_Fiurasek_Hradil_2003}, or alternatively, extremization can
be done by means of numerical up-hill simplex method
\cite{Banaszek_D'Ariano_Paris_Sacchi_2000}.
Any density matrix may be parameterized in diagonal form
(\ref{density_operator}) using  independent (orthogonal) basis
states $|\varphi_k\rangle$ and the variation
(\ref{variation_of_log_likelihood}) may be done along these
rays. Likelihood function depends on the density matrix
through probabilities $p_i$. This yields the  system of
coupled equations $\frac{\delta\,\ln{\mathcal L}(\hat{\rho})}
  {\delta\,\bra\varphi_{k}|} = 0$
for any allowed component $k$. Using the relation
\begin{equation}  \label{state_variation}
  \frac{\delta \, \ln{\mathcal L}(\op{\rho})}
    {\delta \bra \varphi_{k}|} =
    \sum_{i} \frac{f_{i}}{{p}_{i}} \,
    \op{\Pi}_{i} |\varphi_{k}\ket,
\end{equation}
and the normalization ${\rm Tr}[\op{\rho}] = 1$,
the extremal equation
\cite{Hradil_1997,Hradil_Summhammer_Rauch_1999,%
Hradil_Summhammer_2000,Rehacek_Hradil_Jezek_2001,%
Jezek_Fiurasek_Hradil_2003}
for the density operator $\op{\rho}$ reads
\begin{equation}  \label{extremal_operator_equation}
  \op{R} \, \op{\rho} \, = \, \op{\rho}.
\end{equation}
Here
\begin{equation}  \label{extremal_operator_equation_kernel}
  \op{R} =
  \sum_{i} \frac{f_i}{p_i} \, \op{\Pi}_i,
\end{equation}
and probabilities $p_i$ are state dependent
(\ref{flash_in_pixel_probability}). Operator equation
(\ref{extremal_operator_equation}) determines the most
likely solution $\op{\rho}\iest$, for which
$\op{R}(\op{\rho}\iest)=\op{1}$ holds on the Hilbert
space of the state $\op{\rho}\iest$
\cite{Hradil_1997,Rehacek_Hradil_Jezek_2001}.
No prior knowledge about the estimated state is needed.
Results of the measurement itself are sufficient
for analysis.

Let us develop the optical counterpart of this reconstruction
problem. In the spatial domain the extremal equation
(\ref{extremal_operator_equation}) has the form of
{\em integral equation\/} for mutual intensity $\Gamma(x,x')$,
\begin{equation}  \label{extremal_equation}
  \int \! {\rm d}x' \, {\mathcal R}(q,x') \, \Gamma(x',q') =
  \Gamma(q,q'),
\end{equation}
where the resolution of identity
$\op{1} = \int {\rm d}x \; |x\ket \bra x|$
has been used.
Kernel
\begin{equation}  \label{extremal_equation_kernel}
  {\mathcal R}(q,x') = \sum_i \frac{f_i}{p_i} {\mathcal P}_{i}(q,x')
\end{equation}
and functions
\begin{equation}  \label{functional_POVM}
  {\mathcal P}_{i}(q,x') = \int_{\Delta_{i}} \!\! {\rm d}x \,
  h^{\ast}(x,q) \, h(x,x')
\end{equation}
correspond to the operator $\op{R}$ and to the POVM operators
$\op{\Pi}_i$, respectively. The probabilities
(\ref{flash_in_pixel_probability}) of elementary detection
in individual pixels then read
\begin{equation}  \label{explicit_flash_probabilities}
  p_i = \int \!\!\! \int \! {\rm d}q \; {\rm d}q' \,
        \Gamma(q,q') \, {\mathcal P}_{i}(q',q).
\end{equation}
The equation (\ref{extremal_equation}) relates measured
data $f_i$, properties of optical device, and reconstructed
signal $\Gamma(x,x')$.
Dependence on the optical apparatus is expressed via
point-spread function $h(x,x')$ only. However, this mutual
relation is inseparable, since the relation is nonlinear.
In comparison to standard treatments in scalar optics,
no assumptions about statistical nature of the signal have been
done. This seems to be reasonable, since the coherence properties
of the light field may change during the propagation
(van Cittert--Zernike effect \cite{Perina_Coherence_of_Light}).
The proposed formulation anticipates only the knowledge
of the optical apparatus and the measured data without any
prior assumptions about the unknown signal.

Special cases of the generic formulation deserve attention.
Let us assume an iterative solution of the equation
(\ref{extremal_equation}) taking the maximally-ignorant
initial guess represented by the totally mixed uniform state,
$\op{\rho}^{(0)} = \frac1D \op{1}$,
where $1/D$ ensures the trace normalization in $D$-dimensional
Hilbert space. After evaluating the kernel $\op{R}^{(0)}$
we are able to write
down the {\em first iteration\/} of estimated state,
$\op{\rho}^{(1)} = \op{R}^{(0)} \op{\rho}^{(0)} =
\sum_i f_i \op{\Pi}_i / {\rm Tr}[\op{\Pi}_i]$.
In the spatial domain this state has the form of partially
coherent superposition of response functions (\ref{PSF})
weighted by measured data $f_i$, $\sum_i f_i = 1$,
\begin{equation}  \label{first_iteration}
  \Gamma^{(1)}(q,q') = \sum_i f_i \frac
  {\int_{\Delta_{i}} \!\! {\rm d}x \, h^{\ast}(x,q) h(x,q') }
  {\int \! {\rm d}\xi \int_{\Delta_{i}} \!\! {\rm d}x \,
  \left| h(x,\xi) \right|^2 }.
\end{equation}
It is clear that the coherence properties
of estimated signal are changed during repeated iterations
of extremal equation (\ref{extremal_equation}).
Besides the proposed iterative solution
the well-known EM algorithm \cite{Dempster_Laird_Rubin_1977}
completed by unitary step \cite{Rehacek_Hradil_Jezek_2001}
can also be utilized. This guarantees
the convergence and keeps all fundamental
properties (\ref{mutual_intensity_properties})
of partially coherent signal $\Gamma(x,x')$.

As the second special case the {\em totally incoherent
light\/} can be assumed,
\begin{equation}  \label{incoherent_mutual_intensity}
  \Gamma(x,x') = I(x) \delta(x-x'),
\end{equation}
where $\delta(x)$ is Dirac distribution.
The extremal equation then reduces to
\begin{equation}  \label{incoherent_extremal_equation}
  \int \!{\rm d}q' \, {\mathcal R}(q,q') \, I(q') = I(q),
\end{equation}
whereas the probabilities
(\ref{explicit_flash_probabilities}) read
\begin{equation}  \label{incoherent_functional_POVM}
  p_i = \int \!{\rm d}q \, {\mathcal P}_{i}(q,q) \, I(q).
\end{equation}
The relations
(\ref{incoherent_extremal_equation}),
(\ref{extremal_equation_kernel}), and
(\ref{incoherent_functional_POVM}) provide the extremal
equations for unknown optical intensity $I(x)$,
\begin{equation}  \label{Richardson_extremal_equation}
  \sum_i \frac{f_i}{\int\!{\rm d}q\, {\mathcal P}_i(q,q) I(q)}
  \int\!{\rm d}q'\, {\mathcal P}_i(x,q') I(q') = I(x).
\end{equation}
This relationship may be utilized for iterative procedure
as was proposed by Richardson in 1972 for incoherent image
reconstruction. Notice, however, that in the original
derivation \cite{Richardson_1972} the Bayes rule was adopted.
The treatment devised here makes it possible to extend
the solution to the cases of partially or totally
coherent signals.

\section{Numerical example}

In this section we demonstrate the feasibility and advantages
of the presented method by means of the carefully selected
example. The partially coherent testing object is chosen below
the resolution limit of the simple optical device with finite
aperture. Therefore, the observed images do not bear any
resemblance with the true object. In spite of this obvious
limitations the reconstructed object reveals the original
structure. This improvement is achieved by adjusting of the
detector position in transverse and longitudinal directions.
The background noise is added to simulated intensities.
Only these data enter the reconstruction procedure
(\ref{extremal_equation})-(\ref{extremal_equation_kernel}).

Let us consider the optical set-up that
consist of free evolution to the distance $d\iin$,
thin lens with the focal length $f$, and free evolution
to the distance $d\iout$. The lens has the finite aperture
diameter $2a$. The whole device can be transversely 
shifted to a distance $s$ from axial position. The longitudinal
distance $d\iin$ from the object and transverse shift $s$
are parameters of the optical set-up and they may be adjusted
during measurement, while the other parameters are kept
constant. The point-spread function 
$h(x,x_{0}) = h(x,x_{0};d\iin,s) =
\bra x| \op{T}(d\iin,s) |x_{0}\ket$
of the device under consideration reads
\begin{eqnarray}  \label{example_PSF}
  h(x,x_{0}) = \bra x| \,
  \exp{( - {\rm i} \frac{d\iout}{2 k} \op{p}^2 )} \,
  \int_{-a}^{a} \!\! {\rm d}\xi \, |\xi\ket \bra\xi|
  \times & & \nonumber \\
  \times
  \exp{( - {\rm i} \frac{k}{2 f} \op{x}^2 )} \,
  \exp{( - {\rm i} \frac{d\iin}{2 k} \op{p}^2 )} \,
  \exp{( - {\rm i} s \op{p} )} \,
  |x_{0}\ket, & &
\end{eqnarray}
where $k=2\pi/\lambda$ is the longitudinal wave number.
With the help of the relations
$\op{1} = \int {\rm d}x \; |x\ket \bra x|$,
$\op{1} = \int {\rm d}p \; |p\ket \bra p|$, and
$\bra x|p\ket = (2\pi)^{-1/2} \exp{({\rm i}xp)}$
the point-spread function (\ref{example_PSF})
can be rewritten to the form
\begin{equation}  \label{example_PSF_total}
  h(x,x_{0};d\iin,s) = 
  h_{\infty}(x,x_{0}) \, {\cal E}(x,x_{0}).
\end{equation}
Here
\begin{equation}  \label{example_PSF_refract}
  h_{\infty}(x,x_{0}) = {\rm const} \,\,
  e^{ {\rm i} \frac{k}{2} \left[
  \left( \frac{x^2}{d\iout} + \frac{(x_{0}+s)^2}{d\iin} \right)
  - \frac{\theta^2}{\Delta} \right] }
\end{equation}
is the response function of the ideal refractive focusing
device and
\begin{eqnarray}  \label{example_PSF_diffract}
  {\cal E}(x,x_{0}) & = &
  \frac12 \, {\rm erf}\left[
    \frac{1-{\rm i}}{2} \sqrt{k\Delta} \left( \frac{\theta}{\Delta} + a\right)
  \right] - \nonumber \\
  & - &
  \frac12 \, {\rm erf}\left[
    \frac{1-{\rm i}}{2} \sqrt{k\Delta} \left( \frac{\theta}{\Delta} - a\right)
  \right]
\end{eqnarray}
represents the correction to the finite aperture.
This tends to unity for large aperture,
$\lim_{a \rightarrow \infty} {\cal E}(x,x_{0}) = 1$.
The parameter $\Delta$ characterizes
the defocusing from the imaging configuration,
\begin{equation}  \label{Delta_def}
  \Delta = \frac{1}{d\iin} + \frac{1}{d\iout} - \frac{1}{f},
\end{equation}
the parameter $\theta$ is related to the transverse wave number,
\begin{equation}  \label{theta_def}
  \theta = \theta\iin + \theta\iout =
  \frac{x_{0}+s}{d\iin} + \frac{x}{d\iout},
\end{equation}
and the function ${\rm erf}()$ denotes the common error function,
\begin{equation}  \label{erf_def}
  {\rm erf}(z) = \frac{2}{\sqrt{\pi}}
  \int_{0}^{z} \!\! {\rm d}t \, e^{-t^2}.
\end{equation}

\begin{figure}[th]
   \centerline{\scalebox{0.65}{\includegraphics{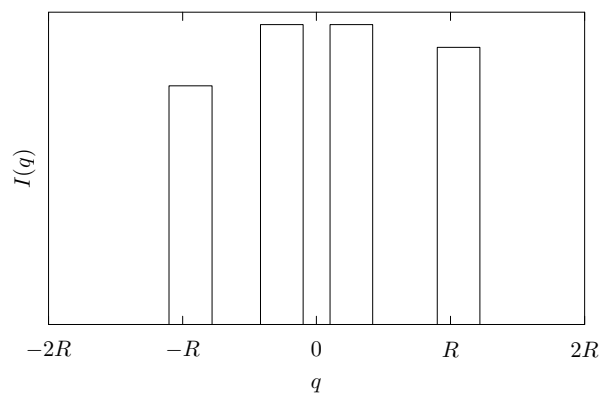}}}
   \caption{Optical intensity $I(q)$ of the true object
            in the input plane.}
   \label{fig_int_true}
\end{figure}

In the present simulation the parameters have been chosen as
$\lambda=600$~nm, $d\iout=1.5$~m, $f=0.5$~m, and $a=0.6$~mm.
The image ($\Delta=0$) appears at the distance $d\iin=0.75$~m.
However, it is blurred due to the small aperture.
In fact, details closer
than the diffraction limit
\begin{equation}  \label{resolution_limit}
  R = {C} \,\, \lambda \frac{d\iout}{a}
\end{equation}
are mapped to two spots with insufficient contrast
according to Rayleigh's criterion.
The factor ${C}$ depends on aperture shape and coherence
properties of the signal. It equals to $0.61$
in the case of circular shape and totally incoherent light,
or to $0.82$ in the case of totally coherent light
(Abbe's resolution limit). We set ${C}=0.5$,
what is equal or smaller than any classical resolution
limit for imaging with partially coherent light.
The testing object consists from four bright spots
separated by dark spaces. The distance $0.15$~mm between
the edges of the central closest spots is $5$-times smaller
than resolution limit $R$ (\ref{resolution_limit}).
The corresponding optical intensity $I(q)$ is shown
in Fig.~\ref{fig_int_true}.
The off-diagonal peaks of mutual intensity $\Gamma(q,q')$
of the testing object
representing the cross-correlations between the spots
are lower than diagonal ones due to the partial coherence.
The contrast
${\cal V} = (I_{\rm max} - I_{\rm min})/(I_{\rm max} + I_{\rm min})$
of the object is ${\cal V} = 1$.

\begin{figure}[th]
   \centerline{\scalebox{0.65}{\includegraphics{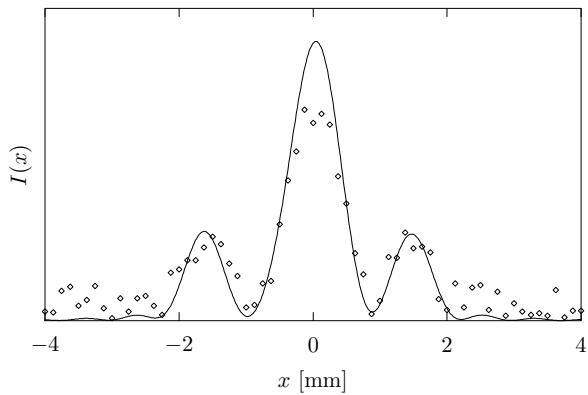}}}
   \caption{Simulated relative frequencies $f_i$ (points)
            affected by $20\%$ of background noise
            sample the optical intensity $I(x)$ in the detection
            plane (lines) for imaging axial arrangement,
            $d\iin=0.75$~m, $s=0$.}
   \label{fig_data}
\end{figure}

\begin{figure}[th]
   \centerline{\scalebox{0.65}{\includegraphics{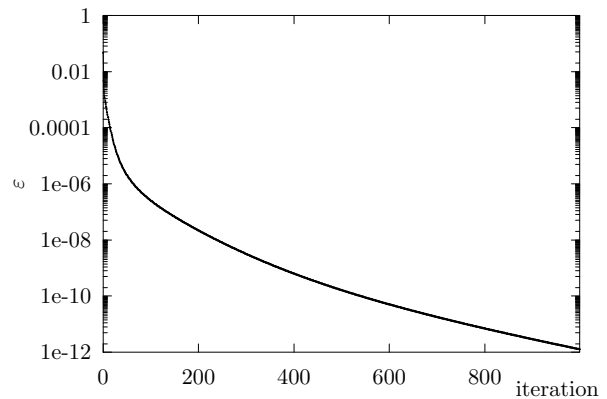}}}
   \caption{The exponentially fast convergence
            of square distance $\varepsilon$
            during the extremization process.}
   \label{fig_converg}
\end{figure}

\begin{figure}[th]
   \centerline{\scalebox{0.65}{\includegraphics{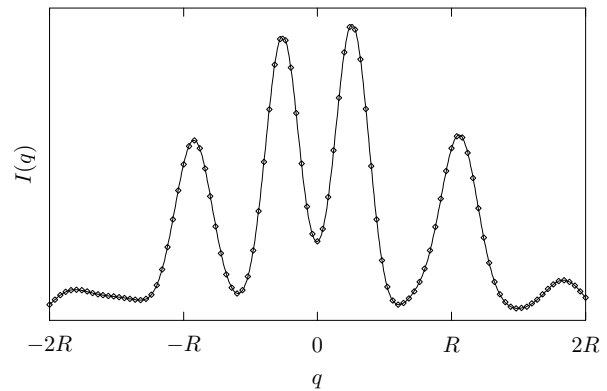}}}
   \caption{Reconstructed optical intensity $I(q)$
            in the input plane.}
   \label{fig_int_est}
\end{figure}

\begin{figure}[th]
   \centerline{\scalebox{1.0}{\includegraphics{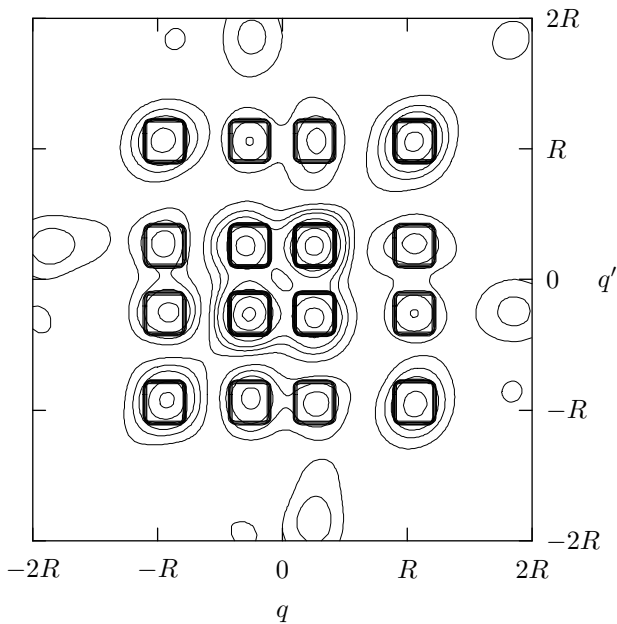}}}
   \caption{Contour lines (thin) of the
            reconstructed mutual intensity $\Gamma(q,q')$.
            The positions of the diagonal bright spots
            as well as the positions of off-diagonal
            correlations match the true ones (thick lines).}
   \label{fig_match}
\end{figure}

The object plane is discretized by $100$ equidistant points
in the interval $[-1.5,1.5]$~mm, or equivalently $[-2R,2R]$.
The corresponding
mutual intensity $\Gamma(q,q')$ is given on the square
mesh of $100 \times 100$ points $(q_m,q'_n)$
in the process of data generation and subsequent
state reconstruction.
Similarly, the detection plane $x$ in the interval $[-4,4]$~mm
is sampled only by $64$ pixels $x_i$
for every longitudinal distance
$d\iin{}_j = (0.75 - 0.05 \, j)$~m, $j=0,\ldots,5$,
and for every transverse shift
$s_l = (-1.2 + 0.3 \, l)$~mm, $l=0,\ldots,8$.
Using the relations
(\ref{functional_POVM}), (\ref{explicit_flash_probabilities}),
and (\ref{example_PSF_total})-(\ref{example_PSF_diffract})
the intensities $p_{ijl} = I_{jl}(x_i)$
in the pixels $i$ can be evaluated for all the 
configurations of the optical set-up. 
For example, in the case of imaging axial configuration
($j=0$, $l=4$) the intensity reveals
the central peak with small side lobes, see Fig.~\ref{fig_data}.
The corresponding under-sampled data $f_{i}$ spoiled by $20\%$
of background noise are shown in the same figure.
These simulated relative frequencies $f_{ijl}$ serve
as an input for the reconstruction procedure
(\ref{extremal_equation})-(\ref{extremal_equation_kernel}).
To solve the extremal equation (\ref{extremal_equation})
we need to find the mutual intensity on the given mesh
as the hermitian matrix of the dimension $100 \times 100$.
As an initial iteration, the uniform incoherent superposition of
all pure states on the supposed space is used. It exhibits
flat intensity profile. It is interesting to note that the final
results seems to be independent of the choice of initial
mutual intensity.
In the course of repeated iterations of the discretized
equation (\ref{extremal_equation}) the difference
$\varepsilon = \int\!\!\int{\rm d}q{\rm d}q'
[ \Gamma^{(n+1)}(q,q')-\Gamma^{(n)}(q,q') ]^2$
between successive iterations can be used as the criterion
for terminating the extremization process.
Numerical results show that the difference $\varepsilon$
reaches the level about $10^{-6}$ after several tens
of iterations and it reaches the level $10^{-12}$ after
approximately $1000$ iterations, see Fig.~\ref{fig_converg}.
The convergence improves slightly for smaller portion
of background noise.
Iterated intensity starts to reveal the four-peak structure
after relatively small number of steps.
The contrast ${\cal V}$ of the central part of the estimated
optical intensity beyond the diffraction limit $R$
reaches the value of $0.56$ after $1000$ iterations,
see Fig.~\ref{fig_int_est}.
The positions and relative intensities of bright spots in estimated
object match very well the structure of the true object.
This is demonstrated in Fig.~\ref{fig_match}.

The numerical simulations clearly show that the proposed
reconstruction algorithm is feasible and could be implemented.
It provides considerable improvement comparing to non-statistical
image processing techniques, and significantly, it yields the
complete information in the form of correlation function.

\section{Conclusion}

The purpose of the presented paper is twofold.
First the tight connection between wave optics and quantum
mechanics has been emphasized. The operator language
routinely used in quantum theory can simplify the
manipulation and description of optical objects, like
partially coherent wave and response function
of optical device. The second goal of the contribution
is the mathematical formulation
of the reconstruction algorithm for partially coherent signal
proceeded from the maximum-likelihood estimation
of mixed quantum state
\cite{Hradil_1997,Hradil_Summhammer_Rauch_1999,%
Hradil_Summhammer_2000,Rehacek_Hradil_Jezek_2001,%
Jezek_Fiurasek_Hradil_2003}.
The solution of extremal equation
by means of repeated iterations has been suggested.
The first iteration has been explicitly
formulated.
The proposed method never yields the non-physical
results.

The feasibility of the method has been verified
by extensive numerical simulations.
The realistic experimental data will be considered
in the forthcoming publication.
The particular numerical example shows the good agreement
between the true and estimated states of partially coherent
light beyond the diffraction limit,
despite of under-sampled data and $20\%$ of background noise.

The method is able to estimate the generic
signal without any prior assumptions,
utilizing only real noisy data. The potential applications
cover wide range of optical inverse problems.
The method may be used for
the state estimation of the localized mode in photonic
band-gap structures (photonic crystals),
for the determination of the near-field short-range
correlation of the signal transmitted through random media
\cite{Emiliani_et_al_2003},
and for the reconstruction of spatial and
coherence properties of light confined and emitted by
modern laser-diode sources.
Moreover, the general quantum origin of the method
allows us to estimate arbitrary continuous (discretized)
partially coherent physical object described by
the correlation function
or the Wigner function.
The reconstruction of de~Broglie wave function of a particle
and the optical homodyne detection of a quantum state
of the light mode are typical examples
\cite{Babichev_et_al_2003}.
In short, the method is applicable to all inverse problems
where the precise knowledge of partially coherent signal
(mixed state) is essential, providing that the measurement
device and real data are known.

\begin{acknowledgments}
  We would like to thank Z. Bouchal, J. Fiur{\'a}{\v{s}}ek,
  M. Du{\v{s}}ek, and J. {\v{R}}eh{\'a}{\v{c}}ek
  for valuable discussions. This work was supported by the EU
  grant under QIPC, project IST-1999-13071 (QUICOV) and by
  Grant LN00A015 of the Czech Ministry of Education.
\end{acknowledgments}


\end{document}